\documentclass{article} 
\usepackage{nips14submit_e,times}
\usepackage{hyperref}
\usepackage{url}
\usepackage{graphicx}
\usepackage{graphicx,etoolbox}

\title{Predict Moves}

\author{
Adam Wang\\
University of California, Los Angeles\\
\And
Steve Chang\\
University of California, Los Angeles\\
\And
John Wilson\\
University of California, Los Angeles\\
}

\nipsfinalcopy

\newif\iffirstimage
\newlength{\imageheight}

\begin{document}

\maketitle

\begin{abstract}
Mobile applications and on-body devices are becoming increasingly ubiquitous tools for physical activity tracking. We propose utilizing a self-tracker's habits to support continuous prediction of whether they will reach their daily step goal, thus enabling a variety of potential persuasive interventions. Our aim is to improve the prediction by leveraging historical data and other qualitative (motivation for using the systems, location, gender) and, quantitative (age) features. We have collected datasets from two activity tracking platforms (Moves and Fitbit) and aim to check if the model we derive from one is generalizable over the other. In the following paper we establish a pipeline for extracting the data and formatting it for modeling. We discuss the approach we took and our findings while selecting the features and classification models for the dataset. We further discuss the notion of generalizability of the model across different types of dataset and the probable inclusion of non standard features to further improve the model's accuracy.
\end{abstract}

\section{Introduction}
People as a whole spend a large amount of time physically inactive. We are inactive while driving, while studying for a big test, while we lounge around in front of the tv after work, many of us work desk jobs, and most likely you are sitting as you read this. More than 50\% of US adults do not reach the recommended amount of physical activity and only 25\% are active at all in their leisure time. 

To meet society's need to become more active, mobile health tracking applications and on-body trackers have flooded the market. One common method for physical activity tracking that many of the current health tracking applications support is step tracking, and it serves as a common denominator that other activities can be translated into. A common recommendation is 10,000 steps per day, or roughly 5 miles, stated as a goal for healthy adults by the American Heart Association and other organizations.

Locke and Latham's goal-setting theory[3] highlights the need to provide feedback on a person progress toward their goal. Pedometers provide feedback by displaying the current step count on devices themselves or in supporting web and smartphone applications; augmented reality applications[4, 5] also help to provide visual feedback of exercise and environment. However, this intervention method places the burden of determining progress on the user, requiring them to look at their tracker several times a day. When surveying the habits of 101 step trackers, we find that trackers create checkpoints throughout their day that they know they need to meet in order to make their goal.

Instead of placing this burden on the step tracker, {\bf what if your pedometer knew at 2pm that you were not going to make your step goal for that day?} It could recommend a slightly different route home from work on public transit that involves more walking, or could suggest a mid-afternoon stroll taking advantage of sunny weather. Our work aims to make timely, actionable feedback possible, exploiting data from previous days to make a prediction.

We begin by conducting a survey of step trackers to gain an understanding of their techniques for discovery and resolution of barriers to reaching their step goal. We utilize data from 80 FitBit wearers to test the feasibility of this prediction, predicting midday step goal achievement between 70-80\% accuracy using a relatively simple decision trees model.  Expanding upon this we gather 79 Moves lifelogger's data, containing information about locations visited as well as steps, develop an advanced feature set, and improve upon midday pedometer prediction accuracy by an average of 5\%. We finally discuss how step goal prediction could encourage step goal achievement in real-world applications, designing four scenarios grounded in Fogg[6] approach to persuasive computing.

For this project we have established a pipeline for extracting the data from their raw form into a database, performing some data transformations in the database and then extracting the data from the database for analysis via various toolkits. We discuss our findings from feature selection and classification model selection and conclude with our future steps.

\section{Pipeline}

\begin{figure}[h]
\begin{center}
\includegraphics*[scale=0.65]{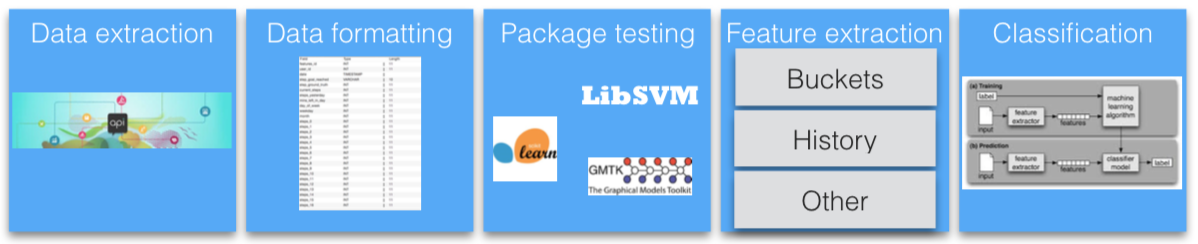}
\end{center}
\caption{Project pipeline : left to right}
\end{figure}

\subsection{Data extraction}

Fitbit data is quite straightforward. For each day we have readings taken at one minute time intervals with each reading being the number of steps taken in that minute. At the end of the day, the final count is compared to the person's daily step goal (10000 steps by default) to see if they met it or not. 

Moves data on the other hand is sporadic. There are two states for a person: Location or Transition. Location is when the person is in a fixed geographical place. Transition is when a person is moving in between two locations. Moves makes a log entry only when there is a change in state. This leads to the data containing only a handful entries per day compared to 1440 rows per day in fitbit. Also, since Moves does not have a default step goal, nearly half the sample does not have a set goal. To this, we will be predicting an estimate of the number of steps the person will achieve at day's end.

The raw data exported from Fitbit and Moves app uses similar JSON format. The app generates a new file with step information for each day. Although we can use similar parsing method to extract raw data, we decided to use different data fields due to the ingrained differences between the app data. One major differentiator was the addition of location data (using Foursquare API) available in the Moves app. We used python scripts to extract the JSON into a MySQL database.

\subsection{Data formatting}

\subsubsection{Fitbit}
The basic data field in Fitbit is steps per minute. However, that seems like more granularity than necessary for our goal. Thus, we use one hour as time resolution to aggregate steps. In addition, we derived some custom fields to aid our process. For example, $steps_.today$ is the number of steps achieved in a specific day; $steps_.yesterday$ is the number of steps achieved in last day; $days_.of_.week$ is a INT from 1-7; $is_.weekday$ is a BOOLEAN value indicates whether today is weekday or weekend.

Most importantly, we created a field $is_.step_.goal_.reached$ as label, calculated by comparing $step_.goal$ and $steps_.today$.

\begin{figure}[h]
\begin{center}
\includegraphics*[scale=0.3]{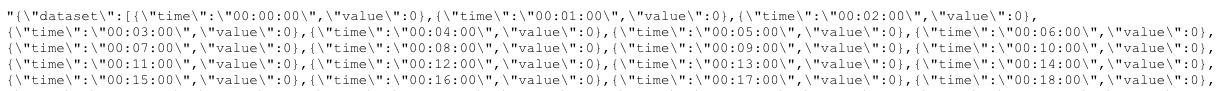}
\end{center}
\label{fig.Fitbit}
\caption{Raw son data as extracted from device}
\end{figure}

\begin{figure}[h]
\begin{center}
\includegraphics*[scale=0.44]{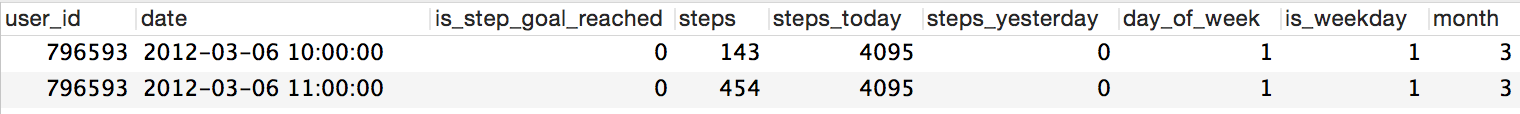}
\end{center}
\label{fig.Fitbit}
\caption{Formatted data with select fields shown}
\end{figure}

\subsubsection{Moves}

The basic data field in Moves is steps in each event (with type location or transition). Each event has start and end time, which is different from fixed Fitbit time interval. It also has meta data for each event, and we can extract location name and type from it. For the sake of generalizability and also to have consistency in the number of features / dimensions, we decided to bucket the daily data into 24 one hour buckets. This basically converted the data from a state transition dataset to a continuous dataset. To achieve this bucketing we divided the number of steps for the state equally among the number of buckets it spanned. While this assumes equal distribution, we believe it is not going to influence the prediction in a negative manner and also the benefit of the dataset now being generalizable across devices is far greater to ignore. In addition, we extract weather and temperature information from Yahoo weather API and will discuss the significance of these features on the classification models.

\begin{figure}[h]
\begin{center}
\includegraphics*[scale=0.3]{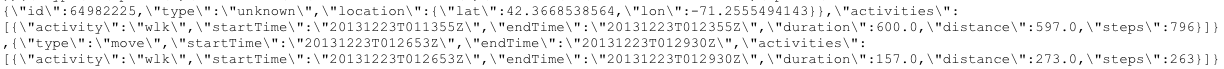}
\end{center}
\label{fig.Fitbit}
\caption{Raw son data as extracted from device}
\end{figure}

\begin{figure}[h]
\begin{center}
\includegraphics*[scale=0.44]{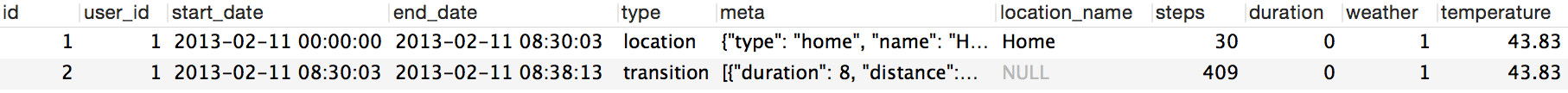}
\end{center}
\caption{Formatted data with select fields shown}
\end{figure}

\subsection{Package testing and Naive analysis}
Instead of starting with random features, we decided to choose the initial features based on insights we got from answering some boring questions. We classify boring questions as those which can be easily answered by running a SQL query over the dataset or running a packaged function over the data. We used Tableau for performing some exploratory visual analytics related to these questions. The main advantage of Tableau is it's power to perform visual analytics - which is quite handy when you are fishing for questions, and not answers, from the data. Example:

Q1: What are the average number of state transitions for a moves user? 
Followup: Do more active people visit more locations? What factors affect the state transition the most?

\begin{figure}[h]
\begin{center}
\includegraphics*[scale=0.2]{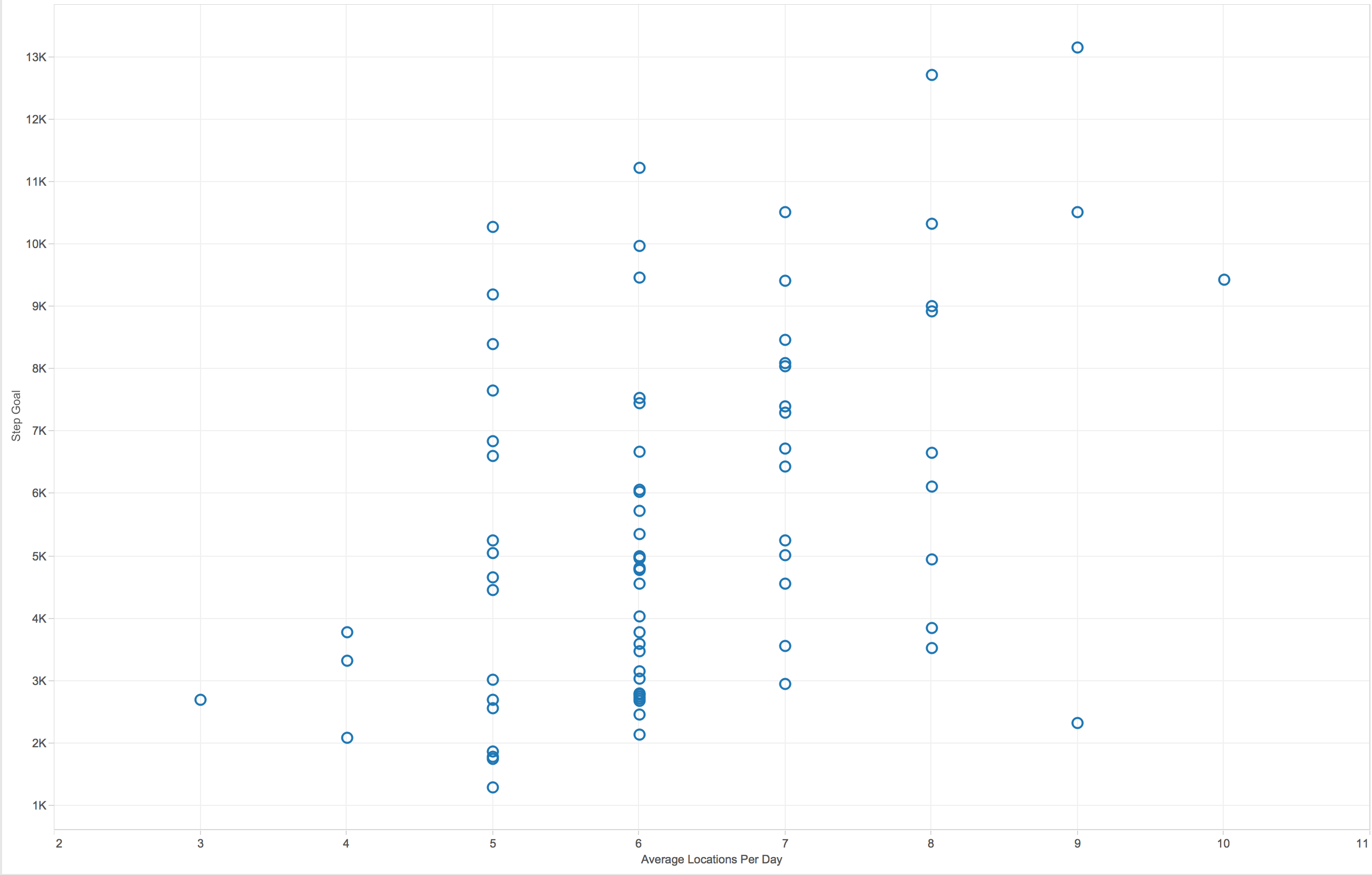}
\includegraphics*[scale=0.2]{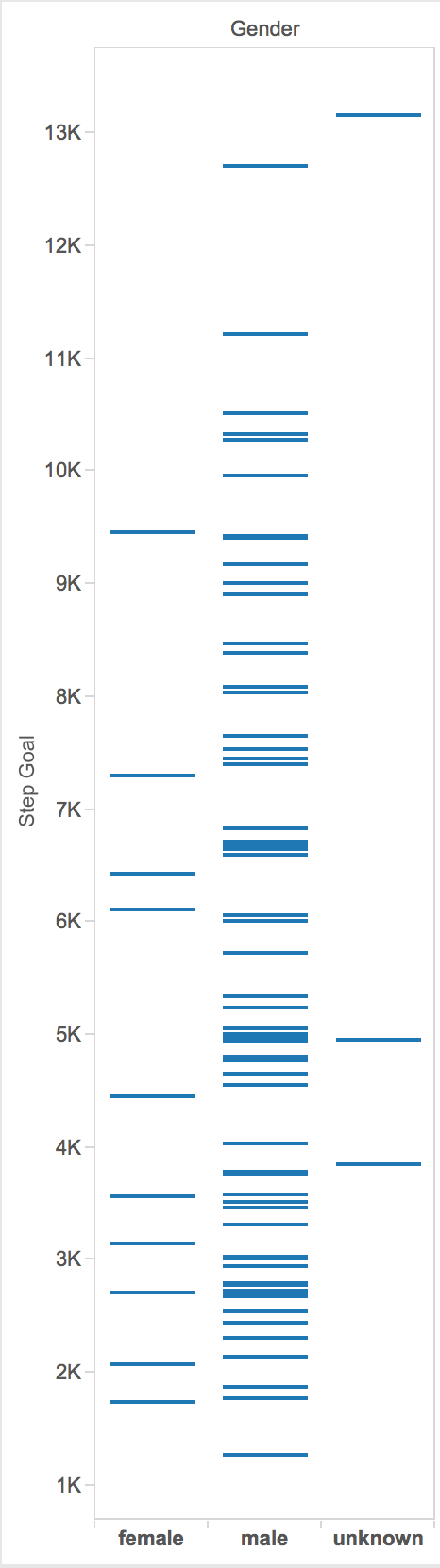}
\end{center}
\caption{Naive Analysis: Average number of locations vs Step goal + Gender vs Step goal}
\end{figure}

\begin{figure}[h]
\begin{center}
\includegraphics*[scale=0.2]{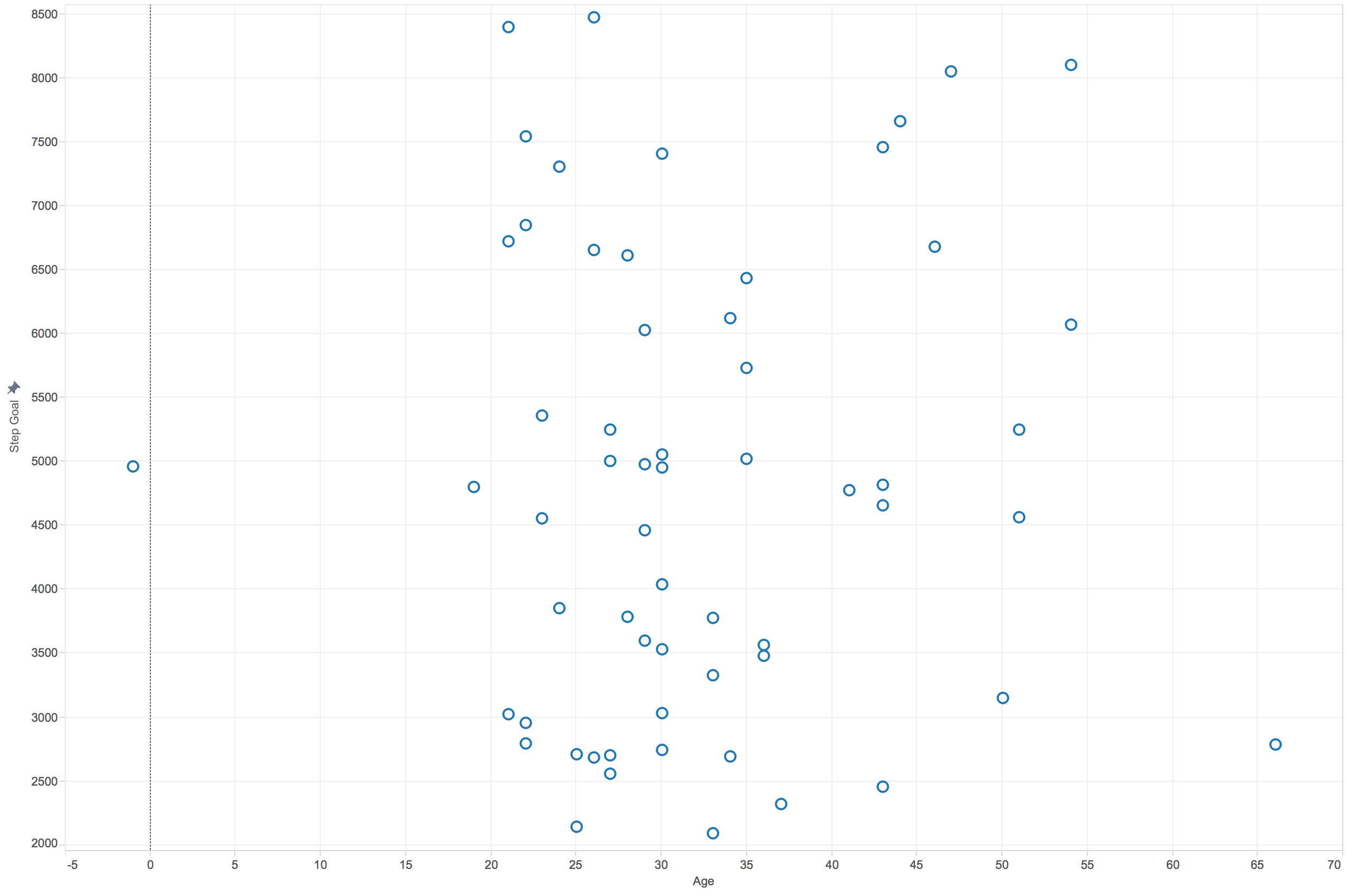}
\end{center}
\caption{Naive Analysis: Age vs Step goal}
\end{figure}

\subsubsection{Scikit[7]}
Scikit is a popular toolkit for data mining and analysis in Python. We used Scikit to better understand our data and do our preliminary feature selection. It helps us understand our data better, in terms of feature selection. We kept the default parameters for each function, so no one will "cheat" to get better performance by tuning parameter. Starting out with Lasso, the assumption of linear data did not achieve a good prediction score (0.164197363261). Following that, in Table 1, we tried Tree-based feature selection[8] to select the most significant features. 
\begin{table}
\label{sample-table}
\begin{center}
\resizebox{\columnwidth}{!}{%
\begin{tabular}{|c|c|c|c|c|c|c|c|c|c|c|c|c|c|c|c|}
Hour & 0 & 1 & 2 & 3 & 4 & 5 & 6 & 7 & 8 & 9 & 10 & 11 & 12 & 13 & 14 \\ [0.5ex]
\hline
11:00 &0.0216 &0.0213 &0.0208 &0.0229 &0.0268 &0.0357 &0.0758 &0.1423 &0.1757 &0.1955 &0.2615 & & & & \\
12:00 &0.0198 &0.0212 &0.0152 &0.0212 &0.0234 &0.0306 &0.0608 &0.1218 &0.1362 &0.1638 &0.1470 &0.2385 & & & \\
13:00 &0.0173 &0.0172 &0.0191 &0.0201 &0.0233 &0.0239 &0.0517 &0.1009 &0.0873 &0.0988 &0.1222 &0.1508 &0.2671 & & \\
14:00 &0.0130 &0.0169 &0.0146 &0.0146 &0.0163 &0.0230 &0.0444 &0.0771 &0.0793 &0.0912 &0.1128 &0.0787 &0.1288 & 0.2890 & \\
15:00 &0.0123 &0.0143 &0.0134 &0.0151 &0.0175 &0.0254 &0.0349 &0.0677 &0.0747 &0.0653 &0.0895 &0.0911 &0.1430 & 0.1383 &0.1968 \\
\end{tabular}%
}
\caption{Tree-based feature selection score}
\end{center}
\end{table}

We found that the steps near current time are more significant, as intuition would suggest. Based on this, we decided to use 4 features (Steps completed in last 4 hours) to build the prediction model for libSVM.

\subsubsection{libSVM[9]}

We build a prediction model each hour, using "Step completed in last 4 hours" as features. Basically, it can predict every hour whether the person hits their daily step goal or not. We use linear kernel for SVM, and parameter C=1. To test our model, we use Five-fold Cross Validation to check prediction scores. In Table 2, the average score shows the accuracy of prediction.

\begin{table}
\label{sample-table}
\begin{center}
\tiny
\begin{tabular}{|c|c|c|c|c|c|c|c|c|c|c|c|c|c|c|c|}
CV score & 1 & 2 & 3 & 4 & 5 & Average\\ [0.5ex]
\hline
11:00 &0.73793103 & 0.64137931 & 0.73793103 & 0.70833333 & 0.69230769 & 70.4\% \\
12:00 &0.75862069 & 0.68275862 & 0.70344828 & 0.73611111 & 0.69930070 & 71.6\%\\
13:00 &0.77931034 & 0.72413793 & 0.72413793 & 0.79166667 & 0.71328671 & 74.7\%\\
14:00 &0.80689655 & 0.76551724 & 0.71034483 & 0.78472222 & 0.72727273 & 75.9\%\\
15:00 &0.82068966 & 0.79310345 & 0.75172414 & 0.84027778 & 0.80419580 & 80.2\% \\
\end{tabular}
\caption{Five-fold Cross Validation score}
\end{center}
\end{table}

\subsubsection{GMTK - Graphical Models Toolkit[10]}
Our initial examination of the data led us to believe that there might be value in looking at graphical modeling techniques for the moves data. We found a toolkit out of UW EE called GMTK. While promising, the learning curve for the toolkit turned out to be quite big for a course project duration. We will look into the toolkit in our future work of the project. 

\subsection{Feature extraction}
We did not find any unexpected results in our naive analysis, which cemented our belief that historical data should be a quite valuable feature to look at. With the bucketing of the Moves data, the tests we run are generalizable to the fitbit dataset too. With that assumption we ran cross validation using various methods listed below to verify the significance of various features. To make sure we are not overfitting and keeping the models generalizable, we did not tweak the default values of the parameters in the models for the time being. The main features we are looking at are the hourly data for the current day. With that said below are the results of our tests:

\subsubsection{LASSO}
For LASSO, the default parameters were  set to Alpha = 1 and Max iterations = 1000.  The table below shows our findings in terms of the coefficient of determination R-squared of the prediction.

\subsubsection{Dimension Reduction: PCA}
For PCA, the default parameter was set to n\_components = 10.  The table below shows our findings in terms of the explained\_variance\_ratio.

Based on these results we can conclude that  the ratio of top 3 explained to total variance is 90\%+, and ratio of top 5 explained to total variance is 99\%+, which indicates that "most of the information" is contained in the first 3 features, and we can pay less attention to features outside of top 5.

\subsubsection{Tree-based feature selection}
For Tree based feature selection, the default parameter was set to n\_estimators = 10.  The table below shows our findings in terms of the feature\_importance.

Based on these results we can conclude that the steps near current time are more significant while the steps completed yesterday, and weekday/weekend have lesser impact.

\begin{figure}[h]
\begin{center}
\includegraphics*[scale=0.5]{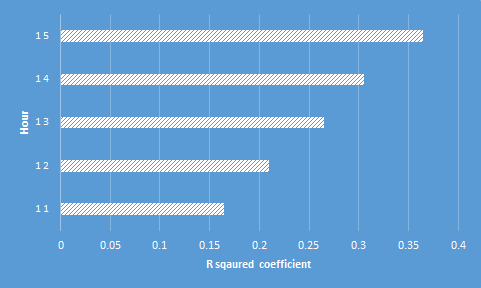}
\includegraphics*[scale=0.5]{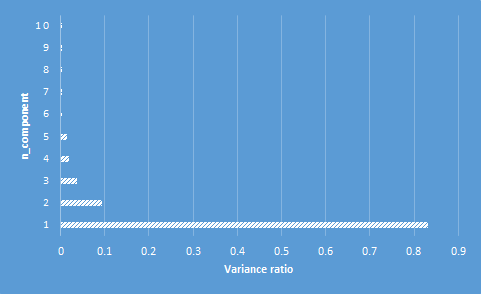}
\includegraphics*[scale=0.5]{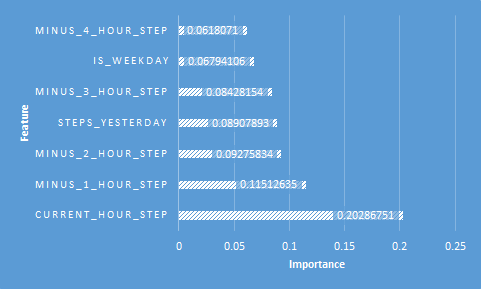}
\end{center}
\caption{Feature selection : graphical comparison of LASSO, PCA. Tree}
\end{figure}

\subsubsection{Weather and Location}
We also did some preliminary testing to check the significance of current weather and location data. We took the tree based feature selection approach and first added the weather parameter to see it's ranking in the list. As we can see the weather parameter is ranked fourth most importance with barely 8.5\% impact on the prediction. Not quite significant.

Similarly, we also tested to see if location (in addition to weather) yielded any significant shift in importance of parameters. But, it just ended up replacing weather as the fourth most significant feature, pushing weather down to sixth. 

\begin{figure}[h]
\begin{center}
\includegraphics*[scale=0.6]{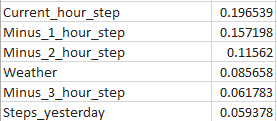}  
\includegraphics*[scale=0.6]{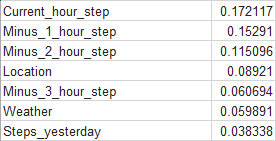}
\end{center}
\caption{Feature selection: Tree based selection with weather and  location}
\end{figure}

Even with these results it would be unwise to disregard the importance of these two features. We will need to spend some time to look at different ways in which we can utilize the location and weather data.

\subsection{Classification}

As we saw during feature extraction, the number of steps taken in the past three one hour windows are the most important features. Based on those results, we used cross validation to get average prediction score in each hour for the various models we discuss below:

\subsubsection{Linear Regression}
Even though ours is a classification problem, we choose to check the accuracy of a linear model in the hope of being able to get a quantified result of prediction i.e. getting an estimate of the number of steps the person will get by the end of the day rather than just checking whether they hit the goal of not. With that in mind we used LASSO for linear regression with the parameters for Alpha = 1 and Max iterations = 1000

We can see the linear regression yielded the worst performance which is not a surprise since regression is not suitable in this classification question and the data is also optimized with classification in mind.

\subsubsection{Nearest Neighbor}
The second approach we tested was nearest neighbor classification. We used Nearest Centroid Classifier for Nearest Neighbor with the metric='euclidean' and shrink\_threshold = None.

The results show that NN has quite good performance with accuracy reaching 80\% around the 2pm mark. However it is interesting to see that the last three scores are same instead of converging to 100\% accuracy. Our understanding is that Nearest Neighbor ignores tiny step differences in last few hours in a day leading to that plateauing.

\subsubsection{SVM}
SVM seemed like the ideal candidate for the problem at hand. We used SVC(Two Classes Classification) with the parameters set to kernel='linear', C=1. 

The results show that SVM is a clear winner albeit NN coming in a close second. As with NN, SVM also hits the 80\% accuracy mark at 2pm and goes on to converge to 100\% at 11 pm. This was the primary component which separated it from the NN and making it the winner.

\begin{figure}[h]
\begin{center}
\includegraphics*[scale=0.55]{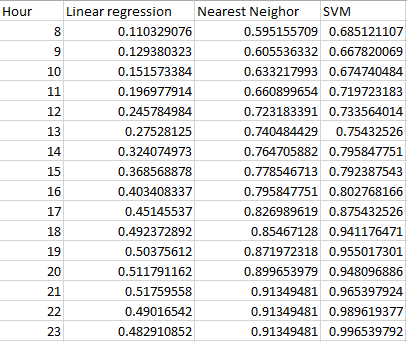}  
\includegraphics*[scale=0.55]{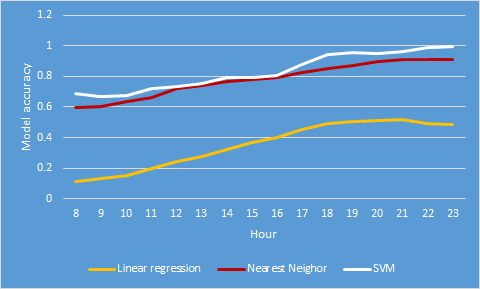}
\end{center}
\caption{Comparison of classification models}
\end{figure}

\subsection{Optimization}

\subsubsection{Nearest Neighbor}

We tried four different distance metrics - euclidean, cosine and manhattan and the results show that euclidean turns out to be the best. For shrinkage threshold the value of each feature for each centroid is divided by the within-class variance of that feature. The feature values are then reduced by shrink\_threshold. This is useful, for removing noisy features.

\begin{figure}[h]
\begin{center}
\includegraphics*[scale=0.5]{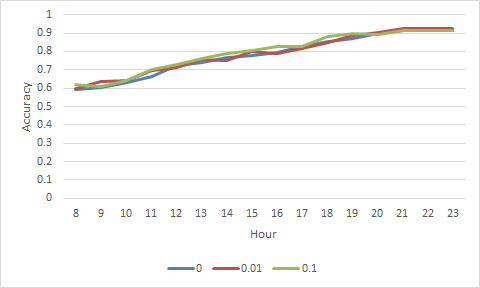}  
\includegraphics*[scale=0.5]{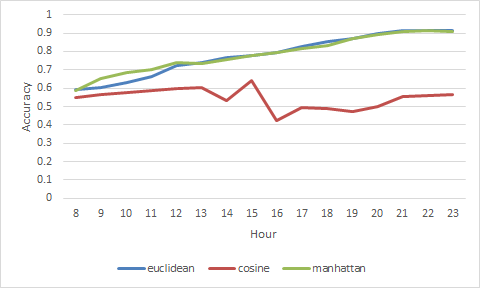}
\end{center}
\caption{Optimization parameters for NN: shrinkage and distance metric  }
\end{figure}

\subsubsection{SVM}

Here C is the penalty parameter of the error term. The optimal values for the various parameters are: C = 0.001, Kernel type = linear and Shrinkage = true (not really impacting the accuracy). 

\begin{figure}[h]
\begin{center}
\includegraphics*[scale=0.5]{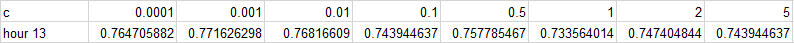}  
\includegraphics*[scale=0.5]{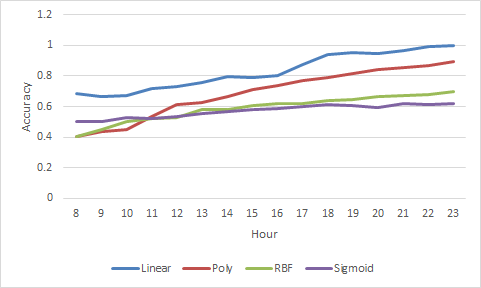}
\includegraphics*[scale=0.5]{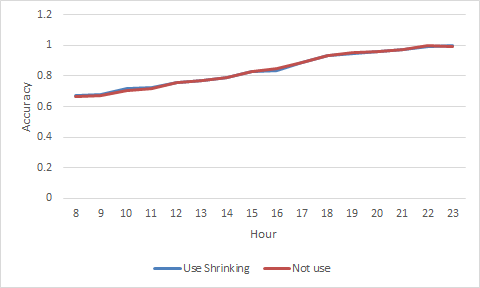}
\end{center}
\caption{Optimizing parameters for SVM: C, type of kernel and shrinkage}
\end{figure}

\section{Conclusion}
\subsection{Discussion}
One of the major primary insight we learned early on was the need to work on features which can be generalized across datasets. This had particular impact on the way we formatted the Moves dataset. We ended up deciding to split a day into (24) one hour buckets. When a state spans across multiple buckets, we just average the number of steps over the buckets. This has no impact whatsoever on the total number of steps a person has taken over the day. The approach will be quite valuable if in the future we decide on adding even more datasets to the pool. 

Secondly, instead of starting with random features, we decided to choose the initial features based on insights we got from answering some boring questions. We classify boring questions as those which can be easily answered by running a SQL query over the dataset or running a packaged function over the data. We used Tableau for performing some exploratory visual analytics related to these questions. While this did not reveal any major insight it was still a handy exercise in terms of understanding the datasets.

Finally, the testing of the various toolkits helped us explore our initial set of features and shape our decisions for the next iterations. It also exposed us to the ever growing landscape of Machine Learning community. We found many things during this phase which we would like to try out once the course if over. 

\subsection{Next steps}
So far, we have explored some initial features and looked at our datasets in a holistic manner. Our next steps include optimizing the feature selection and fine tuning to optimize the existing models we have. We would also like to exploring new models like Markov models and Regression trees to improve our prediction accuracy. We will be taking inspiration from work done in mobility prediction models [11] for tailoring the location feature.
We would also like to steer away from generalizability and see how valuable features like location and weather, ignoring the portability of the model from Moves to Fitbit and just optimizing the model for Moves dataset. This will allow us to use Markov decision models and other interesting approaches which would not be possible on a limiting dataset like Fitbit. Finally, we would like to look at the possibility of grouping users by their activity behavior. The idea here is to change the focus from steps to activity patterns and that too across population instead of just the individual's past.

\subsubsection*{References}

\small{
[1] : https://www.moves-app.com

[2] : http://fitbit.com

[3] : Locke, Edwin A., and Gary P. Latham. A theory of goal setting and task performance. Prentice-Hall, Inc, 1990.

[4] : Zhang, Xiaoyi, Wenyao Xu, Ming-Chun Huang, Navid Amini, and Fengbo Ren. "See UV on your skin: an ultraviolet sensing and visualization system." In Proceedings of the 8th International Conference on Body Area Networks, pp. 22-28. ICST (Institute for Computer Sciences, Social-Informatics and Telecommunications Engineering), 2013.

[5] : Alamri, Atif, Jongeun Cha, and Abdulmotaleb El Saddik. "AR-REHAB: An augmented reality framework for poststroke-patient rehabilitation." IEEE Transactions on Instrumentation and Measurement 59, no. 10 (2010): 2554-2563.

[6] : Fogg, Brian J. "Persuasive computers: perspectives and research directions." In Proceedings of the SIGCHI conference on Human factors in computing systems, pp. 225-232. ACM Press/Addison-Wesley Publishing Co., 1998.

[7] : http://scikit-learn.org

[8] : http://scikit-learn.org/stable/modules/classes.html\#module-sklearn.ensemble

[9] : Chang, Chih-Chung, and Chih-Jen Lin. "LIBSVM: a library for support vector machines." ACM Transactions on Intelligent Systems and Technology (TIST) 2, no. 3 (2011): 27.

[10] : Bilmes, Jeff, and Geoffrey Zweig. "The graphical models toolkit: An open source software system for speech and time-series processing." In Acoustics, Speech, and Signal Processing (ICASSP), 2002 IEEE International Conference on, vol. 4, pp. IV-3916. IEEE, 2002.

[11] : Baratchi, Mitra, Nirvana Meratnia, Paul JM Havinga, Andrew K. Skidmore, and Bert AKG Toxopeus. "A hierarchical hidden semi-Markov model for modeling mobility data." In Proceedings of the 2014 ACM International Joint Conference on Pervasive and Ubiquitous Computing, pp. 401-412. ACM, 2014.
}

\end{document}